# Information Resources Management Framework for Virtual Enterprise

Lingxiang Mao[1,2]

Chery Science Research Institute, Wuhu Anhui, P.R.China[1];
School of Information Management Wuhan University, Wuhan Hubei, P.R.China[2]
maolx@outlook.com

**Abstract:** Virtual enterprise is a new form of organization in recent years which adapt to the IT environment. Information resources management implemented in the virtual enterprise is determined by the form of business organization and information exchange mechanisms. According to the present characteristics of virtual enterprise management, it puts forward the strategies and measures of information resources management framework for virtual enterprise.

## 1. Introduction

With the rapid development of modern information technology, economic globalization, and the coming of knowledge economy, the agile manufacturing, as a representative of the new manufacturing mode has replaced the traditional mass manufacturing mode, and the traditional form of business organization also has a corresponding change. Hence a new type of enterprise organization form -- virtual enterprise or VE for short has been born. The virtual enterprise is the unique organization creation of agile manufacturing, which does not involve the premise of ownership, yet forming a temporary enterprise alliance for a particular task [1]. This kind of dynamic alliance is set up for a special project goal by different types of enterprise can grasp the changing market opportunities in the market with fierce competition. Some scholars believe that the virtual enterprise will become the main forms of enterprise production and operation in the 21st century's market competition [2], which has been research hot spot in modern enterprise management field.

## 2. The role of information resources management in VE

### 2.1. Virtual enterprise overview

In response to the rapidly changing market, to grasp and share more productive resources, and to adapt to the flexible market demand, the companies start to choose cooperation between enterprises. Virtual enterprise is a kind of dynamic alliance which is formed by several enterprise groups based on market opportunities. Davidow and Malone firstly stated virtual enterprise concept in 1992 [3]; Bao and Li consider that virtual enterprise is a complex dynamic combo which use market opportunity as a starting point; use computer and network for the operation platform; use contract for win-win cooperative basis for several independent enterprises with core competences [4]. Therefore, the virtual enterprise is not only a kind of temporary organization form between enterprises, but also is an enterprise effective competition strategy, an organic integration of strategy and structure. Chen, Wang, and Sun has summarized the main characteristics of the virtual enterprise: information technology and industrial information based on the network; exists dynamic based on market opportunity; mutual trust between members of the enterprise; no obvious enterprise boundary; each member enterprise has its own unique core ability, virtual enterprise is a powerful combination; the structure of organization is flattening with flexible organizational structure for easy reconstruction [5].

The virtual enterprise is essentially different aggregation of enterprise core competence, and cooperation is the soul of the virtual enterprise. Its core ability is corresponding to the virtual enterprise core information flow [6]. The virtual enterprise process, such as the production, development, sales and management are virtually. The virtual enterprise can deal with the difficulties and risk increases of enterprise management in the network economy and information technology conditions. Through the cooperation between enterprises, using the advantages of each enterprise resources, it can get effective integration to have fast market reaction ability. The virtual enterprise is different from traditional enterprise's sustainable business model, which is a cycle of the operation, formation and disappeared with the market opportunity. In general, the virtual enterprise members, enterprise in the center position is called "the core enterprise" or "leader", and other enterprises are called "partnership enterprise" or "member".

VE has the superiority and uniqueness which the traditional enterprise does not have. The advantages of VE are that it can quickly respond to market demand, with the aid of the virtual team's strength; it can reduce the cost of research and development; team members can realize the profit and risks sharing.

**2.2. Information resources management overview**

Modern industrial enterprise management environment facing the global economy and the market, the diverse competition situation, the using of information technology, the complanation structure of organization change, information resource management (IRM) is the product of change in the management environment [7]. The modern management thought, mainly evolves through scientific management to humanistic management and to information resources management, just from the management of the things to humanistic management then to information resources as the center in the evolution process.

Information resource management originated in the 1970s, and now it expands from the government documents management field to the enterprise management and scientific research management and so on many domains. In 1985, American scholar F. W. Horton in the book *information resources management* points out, the information resources management is an integrated concept; it sets the different information technologies and fields for an organic whole [8]. C. Wood have argued that the information resources management is an comprehensive of several effective methods in the information management, which combines general management, resource control, computer system management, library management and various policy planning method [9]. J. Huang has pointed out that, information resources management is in order to ensure the effective use of information resources, which is a human management activity using advanced technical means and methods of planning for information resources and related activities, budget, organization, command, control and coordination [6].

IRM definition is not yet unified, but information is strategic weapons that organizations gain a competitive advantage has become a consensus. It creates a strong impact on traditional organization whose management focus transfers from technology to information, and customer choice plays a major role in the IRM. Hence the traditional organization has undergone tremendous changes from form to content. The emergence and development of IRM requires organizational change of the traditional enterprise in turn. Virtual enterprise is the new management model adapted to IRM new organizational forms, which focuses on the strategic role of information resources in the organization and management of virtual enterprise; organization develops to the flat and network; has a high degree of flexibility and learning.

**2.3. IRM supporting role in virtual enterprises**

In the virtual enterprises, "core competencies" for all member companies form the premise of the core competencies of virtual enterprise. Decentralized, dynamic, temporary, instability of virtual enterprises creates a barrier for the play of the core competencies and formation of knowledge networks [10]. This determines the members of the exchange of information in the virtual enterprises are more important than traditional businesses. The writer has researched the information exchange mechanism of the virtual enterprises, and summarizes the mechanism details, which is described by Figure 1 below.

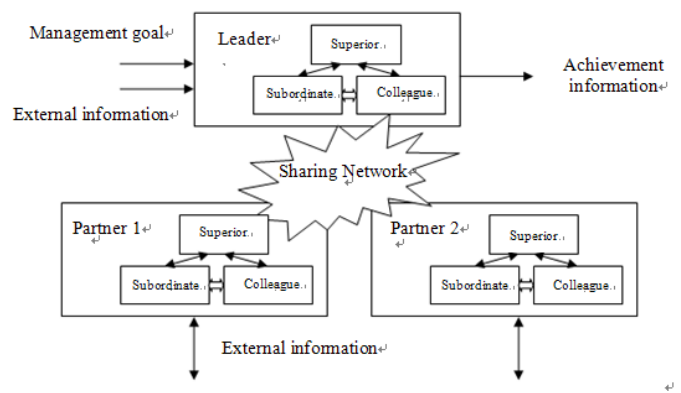

**Figure 1. Information communication mechanism of virtual enterprises**

Through analysis of the mechanisms for information exchange in the virtual enterprise, virtual enterprise IRM has of heterogeneous, autonomous, distribution and so on [11]. The characteristics of the virtual enterprise decided to information resources have an extremely important role in the survival and development of the virtual enterprise. To solve the sharing and management of information resources between members scattered throughout the virtual enterprise, the introduction of IRM thought has great significance.

As it is based on dynamic union of virtual enterprises, communication in the virtual enterprise occupies an important position. In some sense, how to build effective communication between members of virtual enterprise becomes the center of the virtual enterprise management and operation [4]. However, due to the characteristics of virtual enterprise, it is faced with its information communication problems mainly like: different time and space of enterprise makes the physical barrier of virtual enterprise internal members' information communication of dynamic characteristics; emotional disorders of informal communication in enterprise internal members for the dynamic characteristics; internal cultural barriers for the different cultural between member enterprises;

negative problems caused by information technology to form technology barriers of the virtual enterprise internal members' information communication; etc. Therefore IRM has provided a theory and practical base for solving the problems of virtual enterprise communication, and plays an important support role in virtual enterprise.

## 3. The virtual enterprise IRM framework

Before the implementation of virtual enterprise IRM, it is vital to establish a reasonable virtual enterprise IRM operation mechanism which can be called virtual enterprise IRM framework (VEIRMF, for short) that is the foundation and guarantee for information resource management in the virtual enterprise. According to the characteristics of virtual enterprise, combined with the modern management thoughts and information technology, the VEIRMF model has been constructed (as shown in Figure 2), which is constituted by levels of organization management, operation management, data management that is called organization layer, operation layer, data layer respectively. The organization layer supports cooperation and information exchange mechanism for all kinds of members in virtual enterprise; the operation layer supports project management mechanism for the whole life cycle of virtual enterprise; and the data layer provides a foundation of information resource structure for virtual enterprise running.

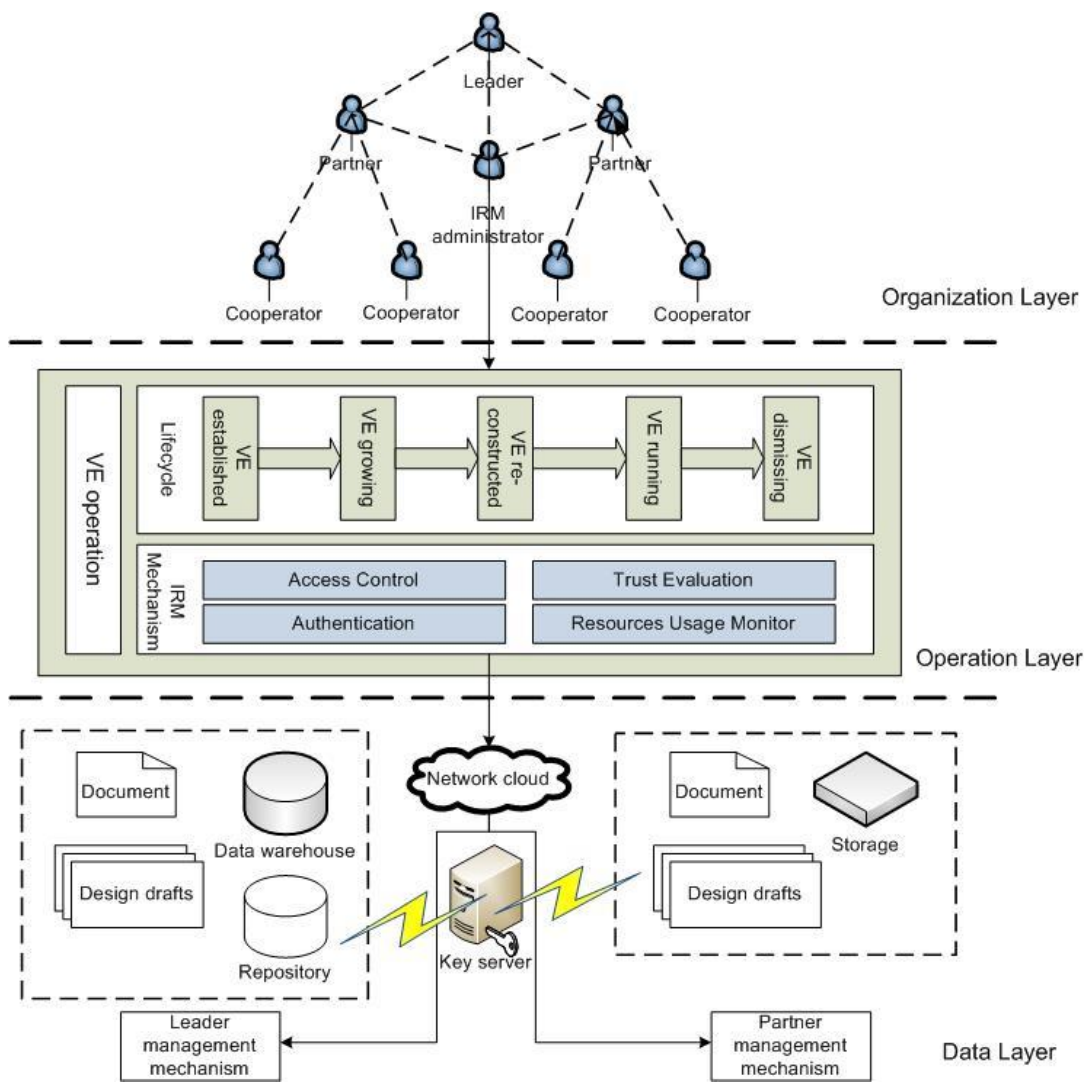

**Figure 2. Virtual enterprise IRM framework model**

### 3.1. Organization layer

The organization layer that is on the top of VEIRMF provides information exchanges and organization cooperation for virtual enterprises members, which include leader, partner, IRM administrator; and other cooperators. The leader should make well relationship with partners related to support resources exchanging with each other, which can promote projects realized.

Sometimes, the partner also needs several cooperators for work coordination and outsourcing, and so cooperator management should also be taken account. IRM administrator is the center of this layer, who can provide indispensable technologies and interactive function.

### 3.2. Operation layer

Operation layer is the realization of the procedure and specific function modules of virtual enterprise information resource management, which can provide the whole life cycle management from formation to demission of the organization. The corresponding IRM function modules related should also be designed, including access control, trust evaluation, authentication and resources usage monitor, etc.

### 3.3. Data layer

Data layer guarantees information resources input, processing, storage and usage in virtual enterprise from material base, which provide proper information management mechanism for leader and partners of the alliance. It uses network clouds and key server to join up the leader and other members of the alliance on the basis of information security. Considering the different information management demand between leader and partner, the data management mechanism should provide corresponding function modules respectively. Hence for leader there should have the functions like documents, data warehouse, design draft, repository and so on; for partner just have documents, storage, design draft.

## 4. The implement of VEIRMF

### 4.1. Define the information resources demand of each member

For VE's characteristics of the geographical distribution, strong life periodicity, flexible and loose form of organization, parallel operation, and heterogeneous members, it decides that virtual enterprise should establish a centralized information resources management mechanism to solve the information communication problems between virtual enterprise members. Because the virtual enterprise management mechanism is different from traditional "mandatory" enterprise, the member enterprises have their autonomy. So the essence of virtual enterprise management is information resource service for the member enterprises. It should fully understand information resource needs of each members of the enterprise to get a good implementation of the virtual enterprise information resources management.

The virtual enterprise is a multi-level complex system, which is made up of many members of the alliance leader, with multiple targets. The information resource demand of virtual enterprise members is decided by its position. The leader mainly focuses on management target and virtual enterprise coordination between members, whose information resources demands mainly have: information of changes in the external environment; daily management information; the goal and the achievements; each member's expertise; enterprise credit information; etc. Because the virtual enterprise is a relatively loose organization, different from the traditional centralized enterprise, it makes every member can be an independent institution to get the external world link. So information resources demands of partners of alliance mainly are described: leader's commands; external environment information; project information; other member's expertise.

### 4.2. IRM department should be established in virtual enterprise

Information resources have become important strategic resources in all kinds of enterprises. In the meantime, organization forms transfer from the traditional "bureaucracy" to the "inverted pyramid" type with the application of information technologies. In this flat organization structure, the communication of information and share are particularly important. It is necessary to establish a professional IRM department, which undertakes core functions and tasks, such as decision, communication, coordination, command, feedback, and research, and integrates information resources of distribute cooperation enterprise, like suppliers, users, agent, research institutions and issue instructions, in an efficient and accurate way to realize the business objectives.

For the virtual enterprise is a dynamic alliance, IRM departments should also be a temporary organization; it is generally run by virtual enterprise information management department of the leader which can bear the corresponding work content.

### 4.3. The content of IRM adjusting to the life cycle of VE

The life cycle of virtual enterprise is the entire course from market opportunities finding, identifying, defining and the virtual enterprise set up, virtual enterprise operation, as well as changes and evolution in market opportunities, to the last dissolved due to the disappearance of market opportunities.

Generally speaking, the virtual enterprise life cycle can be divided into three stages that is the formation, operation, dissolution phases. In the formation stage of the virtual enterprise, the main content of the IRM is: market opportunity discovery, opportunity identification and definition, related enterprises' advantageous resources analysis, virtual organization design, business contracts signed and administration etc.; in the operational phase of the virtual enterprise, IRM includes: information management of VE members, decision support and operational coordination, product definition and manufacturing information, user needs and feedback, real-time operating data management, production experience and skills; in the dissolution phase of the virtual enterprise, the main content of the IRM contains: the completion of all enterprise members, the re-confirmation of enterprise resources of members, communicate contents and approach, customer relationship management, VE experiences and lack analysis.

### 4.4. Support the establishment of the "informal organization" in VE

In the traditional enterprises, the "informal organization" is an important component of the enterprise; it can play the role of the information communication, information sharing, and the formation of corporate culture and heritage, emotional support to employees as a sense of security and a sense of belonging.

However, for the dynamic and shorter life cycle of virtual enterprise, the "informal organization" has not drawn the attention of the virtual enterprise administrators. The writer believes that virtual enterprises still can not ignore the "informal organization". The development of modern information technology makes communication between people greatly expand, in that the social networking services (SNS), instant messaging, mobile phones, weibo and other communication tools can easily communicate, by which people greatly reduce the time and are familiar with the more convenient way of communicating. Therefore to establish "informal organization" platform for the exchange within the virtual enterprise through applications over the network is feasible, but also is necessary. This "informal organization" based network technology the can greatly enhance the sense of belonging of the employees, and enhance the familiarity and trust between employees, and can build a corporate culture that overcoming difficulties together.

### 4.5. Enhance mutual trust between the VE members

Trust is the basis of cooperation, and the core of the transaction. The trust mechanism is the cornerstone of the successful operation of the virtual enterprise [12]. In the virtual enterprise, the "trust" action of member companies is always accompanied by uncertainty and risk, and so the risk can be seen as the core of the virtual enterprise trust. To solve this problem, it is necessary to exchange and communicate among enterprises, reducing uncertainty. Continuous interaction is an important mechanism for members to gather together. Participants are able to identify and develop more fair through the repeated interaction and information exchange, in turn, which enhances the sense of trust.

Virtual enterprise compared to the traditional form of business organization has the characteristics of decentralized and dynamic, which decides more urgent requirements for trust. So we should always put the "service for people" in the first place, to enhance mutual trust between enterprises as the basis to prevent the isolation among people for technologies. Secondly, IRM institutions should also provide more transparent business intelligence for managers for their fair exchange of information, thereby enhancing mutual trust within the virtual enterprise. In addition, the virtual enterprise can use the "relational market" mode and "corporate reputation" mode to establish a relationship of trust between members.

## 5. Conclusion

Modern information technology promoting the changes of the production, forms of enterprise organization and cooperation are also undergoing profound changes. Information resource management this latest management concept has a gradual shift from the traditional enterprise to virtual enterprise. For the characteristics of this particular form of the virtual enterprise organization, the implementation of IRM is very necessary. Considering the characteristics of virtual enterprise, the strategies of VEIRMF implementation have been proposed, through which it realizes and improves the level of IRM in the virtual enterprise. Therefore virtual enterprise and information resource management are complementary: virtual enterprise provides platform for the promotion and development of the IRM, whereas IRM is the inevitable trend of development of virtual enterprise management.

Owning to the new organization model of virtual enterprise, it has not been understood unified, having no uniform implementation standard, which makes virtual enterprise IRM still in the exploration and test stage, with new technologies and new applications emerging continuously. All of these will bring more for the selection and better effect of its implementation, which in the future research work we should keep up with the latest study achievement and timely put it into use in the virtual enterprise information resources management.

## 6. Acknowledgment

Sincerely thanks to Ms Yinghui Sai and Mr Xiaohua Chen, who provide well working circumstance and sufficient research fund for my postdoctoral work.

## References

[1] N Wang. Research on the Strategies of Tacit Knowledge Management for Virtual Enterprises in the Web2. 0 Environment [J]. Information Studies: Theory & Application, 2007, 30(2): 203-205,24.

[2] D G Holt, E P Love, L Heng. The learning organisation: toward a paradigm for mutually beneficial strategic construction alliances [J]. International Journal of Project Management, 2000, 18(6): 415-421.

[3] H W Davidow, M S Malone. The virtual corporation: structuring and revitalizing the corporation for the 21st century [M]. New York: HarperBusiness, 1992.

[4] G Bao, W Li. Communication challenges and countermeasures for virtual enterprises [J]. Science & Technology Progress and Policy, 2005(2): 129-131.

[5] J Cheng, Y Wang, L Sun. The virtual enterprise—enterprise new organization form at the turn of the century [J]. Journal of Industrial Engineering and Engineering Management, 2000(2): 62-64.

[6] J Huang. Research on Integration Management of Enterprise Information Resource [D]. Wuhan: Wuhan University of Technology, 2005: 1-168.

[7] P Wen. Management changes and information resource management [J]. Information Science, 2000(11): 1008-1011.

[8] F W Horton. Information resources management: harnessing information assets for productivity gains in the office, factory, and laboratory [M]. Upper Saddle River: Prentice Hall, 1985.

[9] D Tapscott, A Caston. P Shift: The New Promise of Information Technology[M]. McGraw-Hill Companies, 1992.

[10] Q Ke. Three research perspectives: on virtual enterprise knowledge network [J]. Science and Technology Management Research, 2006(8): 197-198,20.

[11] Y Dou, P Zhao. Information infrastructure research within virtual enterprise [J]. Information Science, 2004(4): 400-402.

[12] X Qiao. The virtual enterprise trust mechanism research [D]. Chongqing: Chongqing University, 2004: 1-70.